# What will computational modelling approaches have to say in the era of atomistic cryo-EM data?

James S. Fraser[1], Kresten Lindorff-Larsen[2], Massimiliano Bonomi[3,*]

[1]Department of Bioengineering and Therapeutic Sciences, University of California, San Francisco, San Francisco, CA, USA.

[2]Structural Biology and NMR Laboratory, Linderstrøm-Lang Centre for Protein Science, Department of Biology, University of Copenhagen, Copenhagen, Denmark.

[3]Structural Bioinformatics Unit, Department of Structural Biology and Chemistry; CNRS UMR 3528; C3BI, CNRS USR 3756; Institut Pasteur, Paris, France.

*To whom correspondence should be addressed: mbonomi@pasteur.fr

## Abstract

The focus of this viewpoint is to identify, in the era of atomistic resolution cryo-electron microscopy data, the areas in which computational modelling and molecular simulations will bring valuable contributions to structural biologists and to give an overview of some of the existing efforts in this direction.

Over the last decade, cryo-electron microscopy (cryo-EM) has become an invaluable technique in structural biology. Thanks to the recent developments in instrumentation, sample preparation, and image-processing software, cryo-EM has now reached atomistic resolution (a resolution high enough to allow structural modelling of unique positions for most atoms in a protein). The progress has been rapid: of the 1818 single-particle EM maps with resolution better than 15 Å deposited during the course of 2019 in the EMDB database, 61% and 86% were resolved at resolution better than 4 Å and 6 Å, respectively. The resolution record is currently held by a 1.54 Å resolution structure of apoferritin (EMD-9865). As of today, 45% of all the single-particle EM maps deposited over the years reported a resolution better than 5 Å[*].

As the number of available atomistic cryo-EM datasets rapidly increases, one can wonder what computational modelling and molecular simulations will bring to the table. The focus of this brief viewpoint is to identify the areas in which these computational approaches can complement the vast amount of information provided by cryo-EM data. Simulation approaches have traditionally been leveraged to derive single structures that are good fits to the density maps.[1-3] Here, we highlight some of the new efforts to expand this direction, while acknowledging that more exhaustive overview of traditional modeling and validation methods have recently been published.[4, 5]

Cryo-EM has uses beyond being a powerful technique for determining PDB structures. Due to the single molecule nature of the experiment, cryo-EM presents new opportunities to study the conformational landscape of dynamic macromolecular systems. Such characterizations can be directly obtained from the raw data, *i.e.* the set of single-particle, two-dimensional (2D) images of the system deposited on a thin layer of vitreous ice, a relatively natural environment. While there have been initial attempts to define conformational landscapes[6-8], currently the most common practice is to derive multiple three-dimensional (3D) density maps in distinct conformational or compositional states. Distinct atomistic models can then be built into these maps and some information about their relative populations can be obtained from the number of particles underlying their reconstruction.

---

[*] Source https://www.ebi.ac.uk/pdbe/emdb/statistics_num_res.html/; last accessed 1/22/2020

Ensemble modeling[9] may be able to bridge the current gap between the long term ideal of "conformational landscapes" and the current reality of "single structures". Computational advances will play a key role in this new direction. While image classification techniques are often able to distinguish distinct conformational states at the level of 2D class averages, highly-dynamic parts of the system are sometimes difficult to identify even with focused classification approaches.[10] Low resolution regions of a cryo-EM map might therefore hide multiple different, but modellable, conformations whose densities have been averaged out in the processing of the raw data.

Cases where these regions exhibit continuous dynamics are particularly challenging. Highly-flexible parts of otherwise well-ordered systems, such as short loops or other disordered regions, are hard to resolve by cryo-EM image-processing alone, yet they are often crucial for specific biological functions. In these cases, traditional modelling approaches that yield a single structure or multiple independently refined models[11] into a density map may not be helpful, as they may not faithfully represent the underlying dynamics of the system.

Recently, several different computational approaches aimed at determining conformational ensembles consistent with ensemble-averaged experimental data have been developed.[12, 13] These methods have traditionally been used in combination with solution experiments, such as Nuclear Magnetic Resonance (NMR) spectroscopy or small-angle X-ray scattering, either on-the-fly during a molecular dynamics (MD) simulation to improve the quality of the underlying force field or *a posteriori* to refine an ensemble previously generated using MD or other modelling techniques. These methods can now be extended to generate structural ensembles from cryo-EM density maps.

Metainference[14] is an integrative method for determining structural ensembles using noisy, ensemble-averaged data based on a Bayesian probabilistic framework. In this approach, an ensemble is generated by a multi-replica MD simulation guided by a hybrid energy function that combines the cryo-EM density map, which is converted into spatial restraints, with a more traditional force field. Molecular mechanics force fields used in MD simulations are becoming more and more accurate in the description of different environments and their interaction with macromolecules. Therefore, there is an emerging opportunity for integrative methods that combine cryo-EM data with MD to provide a more accurate description of the interactions between macromolecules and other, smaller, components (lipids, ions, solvent, ligands, etc) as they become visible in atomistic maps. Although these approaches usually come at a higher computational cost compared to standard real-space refinement techniques, they can provide crucial insights into the interaction of proteins with their environment. Metainference is a powerful method for characterizing the conformational heterogeneity in maps that resist further 3D classification. It has been used recently to determine the flexibility of the extracellular domain of the membrane receptor STRA6[15], the dynamics of the N-terminal gating region of the ClpP protease[16] and the effect of acetylation of residue K40 on α-tubulin.[17]

As the resolution of cryo-EM maps increases, more components of the system are becoming more and more visible at an unprecedented level of detail, such as ordered water molecules, lipids, and ions. State-of-the-art software for single-structure refinement typically provide either none or simplified physico-chemical models of the environment surrounding biological systems. For example, soluble proteins are traditionally refined in 3D density maps using energy functions that describe only basic stereochemical properties and not the surrounding environment, neither using an implicit nor explicit water model. Furthermore, even for the ensemble-modelling approaches that use more accurate molecular mechanics force fields, such as metainference, modelling of ordered water and lipid densities is still challenging and will require further methodological developments.

One of the major challenges for ensemble-modelling approaches is the ability to distinguish conformational heterogeneity from noise in the data. Both these causes can result in the presence of regions at lower resolution in atomistic maps. To overcome this challenge, a modelling approach that accounts for the simultaneous presence of both structural heterogeneity and noise is required along with *i)* structural priors that can

(alone) describe the dynamics of a system sufficiently well and *ii)* accurate estimates of the experimental errors at play.

Rather than relying on 3D maps, new approaches are emerging that use the raw 2D particle stacks, which are sometimes available in the public EMPIAR database[†].[18] Two notable examples are manifold embedding[6] and BioEM.[19] Manifold embedding projects the entire sampled continuum of states in a coordinate system tailored to the cryo-EM images. BioEM uses *i)* a probabilistic Bayesian framework to assess the consistency between a structural model and the set of single-particle cryo-EM images and *ii)* a maximum-parsimony approach to identify a minimal ensemble that can collectively explain the data. New approaches are continually emerging, such as using a variational autoencoder to connect unlabelled 2D cryo-EM images with continuous distributions over 3D densities.[20]

The major advantage of these approaches consists in using the raw data prior to any clustering or averaging procedure, therefore fully embodying the single-molecule nature of the cryo-EM experiment. Currently methods development in this area is limited by the sporadic practice of depositing raw data in the EMPIAR database. These methods are mostly limited by the low signal-to-noise ratio of individual particles, which will be mitigated in time as detectors continue to develop.

Computational modelling can also provide information on several aspects of the cryo-EM experiment that are needed to relate the results to the solution, room temperature ensemble. For example, how particles interact with the air-water interface[21-23] could be studied by multiscale methods and tap into the rich history of using simulations for studying unfolding. Simulations could also be leveraged to determine the vitrification process on the effective "temperature" of the resulting ensemble of molecules. During the cryo-EM experiment, prior to data collection, the sample is prepared in solution at room temperature and then rapidly cooled down to cryogenic temperatures. The timescale of freezing is not fully known, but may take from hundreds of microseconds to a few milliseconds. In this timescale, scarcely populated "excited" states might collapse in neighboring free-energy minima, while more populated, stable states should be less affected. On a more local scale, rotamers and loops are often highly mobile on the microsecond timescale, and thus they may very well have time to structurally reorganize during freezing. Therefore, the conformational landscape represented by the cryo-EM single-particle images might differ in subtle, but potentially important ways from the room temperature, biologically relevant ensemble.

---

[†] https://www.ebi.ac.uk/pdbe/emdb/empiar

One potential approach to study these effects is through non-equilibrium MD. By simulating the freezing process starting from a set of conformers extracted from a solution, room-temperature ensemble, which can be modelled using a long equilibrium MD simulation at 300K, down to the cryogenics ensemble, such simulations can highlight the potential differences between the cryo-EM and the room temperature ensembles. To validate this procedure, one can calculate the agreement of the ensemble with the set of single-particle 2D images prior and after the simulated freezing process and assess whether the cryo-EM ensemble better fits the data. Recent experiments have begun to address these questions experimentally by incubating the samples at different temperatures prior to the freezing and vitrification process[24, 25], and will provide useful points of comparison for molecular simulations.

While the rate of deposition of atomistic cryo-EM maps in the EMDB database is rapidly increasing, a substantial part of the available data is still at medium and low resolution. In these situations, integrative modelling approaches aimed at optimally combining different types of experiments provide an excellent way to complement the scarce information content of medium-low resolution cryo-EM data and thus to enable determining more accurate and precise single-structure models. A recent successful example of combination of cryo-EM with NMR data is the determination of the structure of the 468 kDa large dodecameric aminopeptidase TET2 to a precision below 1 Å starting from a 4.1 Å resolution EM map.[26] At this resolution, it was difficult to trace the backbone and assign the sequence using the cryo-EM data alone, but by combining with secondary structures modelled using NMR data it was possible to determine precise models using both the original data and data artificially truncated to 8 Å.

Rather intriguing is also the possibility of using integrative ensemble-modelling approaches to combine cryo-EM data with other experiments to obtain more accurate protein conformational ensembles. For example, one could envision incorporating NMR ensemble-averaged data to improve the characterization of highly-flexible parts of biological systems that are often averaged out in the cryo-EM classification and reconstruction processes.

The major challenge in single-structure and ensemble integrative approaches is how to balance the information provided by different types of experimental data. In these regards, Bayesian statistics[27, 28] is an effective framework that can be used to combine all sources of information available on the system, being experimental data or physico-chemical knowledge, by weighting them based on accuracy and information content.

In conclusion, while we are in the middle of an explosion of the number of atomistic cryo-EM data available, computational modelling and molecular simulations can still play an important role. These methods will certainly provide in the future essential contributions in many areas of structural biology, from improving the description of protein conformational ensembles, to elucidating the effect of freezing on the behavior of biological systems, accurately characterizing complex physico-chemical environments, and integrating cryo-EM with other types of experimental data.

**References**

1. Kidmose, R. T.; Juhl, J.; Nissen, P.; Boesen, T.; Karlsen, J. L.; Pedersen, B. P., Namdinator - automatic molecular dynamics flexible fitting of structural models into cryo-EM and crystallography experimental maps. *IUCrJ* **2019**, 6, 526-531.
2. Igaev, M.; Kutzner, C.; Bock, L. V.; Vaiana, A. C.; Grubmuller, H., Automated cryo-EM structure refinement using correlation-driven molecular dynamics. *Elife* **2019**, 8, e43542.
3. Singharoy, A.; Teo, I.; McGreevy, R.; Stone, J. E.; Zhao, J. H.; Schulten, K., Molecular dynamics-based model refinement and validation for sub-5 angstrom cryo-electron microscopy maps. *Elife* **2016**, 5, e16105.
4. Malhotra, S.; Trager, S.; Dal Peraro, M.; Topf, M., Modelling structures in cryo-EM maps. *Curr. Opin. Struct. Biol.* **2019**, 58, 105-114.
5. Lopez-Blanco, J. R.; Chacon, P., Structural modeling from electron microscopy data. *Wiley Interdiscip. Rev. Comput. Mol. Sci.* **2015**, 5, 62-81.
6. Dashti, A.; Schwander, P.; Langlois, R.; Fung, R.; Li, W.; Hosseinizadeh, A.; Liao, H. Y.; Pallesen, J.; Sharma, G.; Stupina, V. A.; Simon, A. E.; Dinman, J. D.; Frank, J.; Ourmazd, A., Trajectories of the ribosome as a Brownian nanomachine. *Proc. Natl. Acad. Sci. USA* **2014**, 111, 17492-17497.
7. Haselbach, D.; Komarov, I.; Agafonov, D. E.; Hartmuth, K.; Graf, B.; Dybkov, O.; Urlaub, H.; Kastner, B.; Luhrmann, R.; Stark, H., Structure and Conformational Dynamics of the Human Spliceosomal B(act) Complex. *Cell* **2018**, 172, 454-464.e11.
8. Lu, Y.; Wu, J.; Dong, Y.; Chen, S.; Sun, S.; Ma, Y. B.; Ouyang, Q.; Finley, D.; Kirschner, M. W.; Mao, Y., Conformational Landscape of the p28-Bound Human Proteasome Regulatory Particle. *Mol. Cell.* **2017**, 67, 322-333.e6.
9. Bonomi, M.; Vendruscolo, M., Determination of protein structural ensembles using cryo-electron microscopy. *Curr. Opin. Struct. Biol.* **2019**, 56, 37-45.
10. Scheres, S. H. W., Processing of Structurally Heterogeneous Cryo-EM Data in RELION. *Methods Enzymol.* **2016**, 579, 125-157.
11. Herzik, M. A., Jr.; Fraser, J. S.; Lander, G. C., A Multi-model Approach to Assessing Local and Global Cryo-EM Map Quality. *Structure* **2019**, 27, 344-358.e3.

12. Bonomi, M.; Heller, G. T.; Camilloni, C.; Vendruscolo, M., Principles of protein structural ensemble determination. *Curr. Opin. Struct. Biol.* **2017**, 42, 106-116.

13. Bottaro, S.; Lindorff-Larsen, K., Biophysical experiments and biomolecular simulations: A perfect match? *Science* **2018**, 361, 355-360.

14. Bonomi, M.; Camilloni, C.; Cavalli, A.; Vendruscolo, M., Metainference: A Bayesian inference method for heterogeneous systems. *Sci. Adv.* **2016**, 2, e1501177.

15. Bonomi, M.; Pellarin, R.; Vendruscolo, M., Simultaneous Determination of Protein Structure and Dynamics Using Cryo-Electron Microscopy. *Biophys. J.* **2018**, 114, 1604-1613.

16. Vahidi, S.; Ripstein, Z. A.; Bonomi, M.; Yuwen, T.; Mabanglo, M. F.; Juravsky, J. B.; Rizzolo, K.; Velyvis, A.; Houry, W. A.; Vendruscolo, M.; Rubinstein, J. L.; Kay, L. E., Reversible inhibition of the ClpP protease via an N-terminal conformational switch. *Proc. Natl. Acad. Sci. USA* **2018**, 115, E6447-E6456.

17. Eshun-Wilson, L.; Zhang, R.; Portran, D.; Nachury, M. V.; Toso, D. B.; Lohr, T.; Vendruscolo, M.; Bonomi, M.; Fraser, J. S.; Nogales, E., Effects of alpha-tubulin acetylation on microtubule structure and stability. *Proc. Natl. Acad. Sci. USA* **2019**, 116, 10366-10371.

18. Iudin, A.; Korir, P. K.; Salavert-Torres, J.; Kleywegt, G. J.; Patwardhan, A., EMPIAR: a public archive for raw electron microscopy image data. *Nat. Methods* **2016**, 13, 387-388.

19. Cossio, P.; Hummer, G., Bayesian analysis of individual electron microscopy images: Towards structures of dynamic and heterogeneous biomolecular assemblies. *J. Struct. Biol.* **2013**, 184, 427-437.

20. Zhong, E. D.; Bepler, T.; Davis, J. H.; Berger, B., Reconstructing continuous distributions of 3D protein structure from cryo-EM images. *arXiv* **2019**, 1909.05215.

21. Noble, A. J.; Wei, H.; Dandey, V. P.; Zhang, Z.; Tan, Y. Z.; Potter, C. S.; Carragher, B., Reducing effects of particle adsorption to the air-water interface in cryo-EM. *Nat. Methods* **2018**, 15, 793-795.

22. Noble, A. J.; Dandey, V. P.; Wei, H.; Braschi, J.; Chase, J.; Acharya, P.; Tan, Y. Z.; Zhang, Z. N.; Kim, L. Y.; Scapin, G.; Rapp, M.; Eng, E. T.; Rice, W. J.; Cheng, A. C.; Negro, C. J.; Shapiro, L.; Kwong, P. D.; Jeruzalmi, D.; des Georges, A.; Potter, C. S.;

Carragher, B., Routine single particle cryoEM sample and grid characterization by tomography. *Elife* **2018**, 7, e34257.

23. D'Imprima, E.; Floris, D.; Joppe, M.; Sanchez, R.; Grininger, M.; Kuhlbrandt, W., Protein denaturation at the air-water interface and how to prevent it. *Elife* **2019**, 8, e42747.

24. Chen, C. Y.; Chang, Y. C.; Lin, B. L.; Huang, C. H.; Tsai, M. D., Temperature-Resolved Cryo-EM Uncovers Structural Bases of Temperature-Dependent Enzyme Functions. *J. Am. Chem. Soc.* **2019**, 141, 19983-19987.

25. Singh, A. K.; McGoldrick, L. L.; Demirkhanyan, L.; Leslie, M.; Zakharian, E.; Sobolevsky, A. I., Structural basis of temperature sensation by the TRP channel TRPV3. *Nat. Struct. Mol. Biol.* **2019**, 26, 994-998.

26. Gauto, D. F.; Estrozi, L. F.; Schwieters, C. D.; Effantin, G.; Macek, P.; Sounier, R.; Sivertsen, A. C.; Schmidt, E.; Kerfah, R.; Mas, G.; Colletier, J. P.; Guntert, P.; Favier, A.; Schoehn, G.; Schanda, P.; Boisbouvier, J., Integrated NMR and cryo-EM atomic-resolution structure determination of a half-megadalton enzyme complex. *Nat. Commun.* **2019**, 10, 2697.

27. Rieping, W.; Habeck, M.; Nilges, M., Inferential structure determination. *Science* **2005**, 309, 303-306.

28. Orioli, S.; Larsen, A. H.; Bottaro, S.; Lindorff-Larsen, K., How to learn from inconsistencies: Integrating molecular simulations with experimental data. *arXiv* **2019**, 1909.06780.